\documentclass[sigconf]{acmart}

\usepackage{subfigure}
\usepackage{booktabs} 
\usepackage{multirow}
\usepackage{url}
\usepackage{makecell}
\setcopyright{acmlicensed}

\newcommand{\specialcell}[2][c]{%
  \begin{tabular}[#1]{@{}c@{}}#2\end{tabular}}

\copyrightyear{2018} 
\acmYear{2018} 
\setcopyright{acmlicensed}
\acmConference[WebSci'18]{10th ACM Conference on Web Science}{May 27--30, 2018}{Amsterdam, The Netherlands}
\acmBooktitle{Proceedings of the 10th ACM Conference on Web Science}
\acmPrice{15.00}
\acmDOI{10.1145/3201064.3201081}
\acmISBN{978-1-4503-5563-6/18/05}

\begin{document}
\title[Analyzing Right-wing YouTube Channels: Hate, Violence and Discrimination]{Analyzing Right-wing YouTube Channels:\\Hate, Violence and Discrimination}
\author{Raphael Ottoni}
\affiliation{%
  \institution{DCC/UFMG, Brazil}
}
\email{rapha@dcc.ufmg.br}

\author{Evandro Cunha}
\affiliation{%
  \institution{DCC/UFMG, Brazil}
  \institution{LUCL/Univ. Leiden, The Netherlands}
}
\email{evandrocunha@dcc.ufmg.br}

\author{Gabriel Magno}
\affiliation{%
  \institution{DCC/UFMG, Brazil}
}
\email{magno@dcc.ufmg.br}

\author{Pedro Bernardina}
\affiliation{%
  \institution{DCC/UFMG, Brazil}
}
\email{pedronascimento@dcc.ufmg.br}

\author{Wagner Meira Jr.}
\affiliation{%
  \institution{DCC/UFMG, Brazil}
}
\email{meira@dcc.ufmg.br}

\author{Virgilio Almeida}
\affiliation{%
  \institution{DCC/UFMG, Brazil}
  \institution{Berkman Klein Center/Harvard, USA}
}
\email{virgilio@dcc.ufmg.br}

\renewcommand{\shortauthors}{Ottoni, Cunha,  Magno, Bernardina, Meira Jr., and Almeida}

\begin{abstract}
As of 2018, YouTube, the major online video sharing website, hosts multiple channels promoting right-wing content.
In this paper, we observe issues related to hate, violence and discriminatory bias in a dataset containing more than 7,000 videos and 17 million comments.
We investigate similarities and differences between users' comments and video content in a selection of right-wing channels and compare it to a baseline set using a three-layered approach, in which we analyze (a) lexicon, (b) topics and (c) implicit biases present in the texts.
Among other results, our analyses show that right-wing channels tend to (a) contain a higher degree of words from ``negative'' semantic fields, 
(b) raise more topics related to war and terrorism, and
(c) demonstrate more discriminatory bias against Muslims (in videos) and towards LGBT people (in comments).
Our findings shed light not only into the collective conduct of the YouTube community promoting and consuming right-wing content, but also into the general behavior of YouTube users.
\end{abstract}

\ccsdesc[500]{Human-centered computing~Empirical studies in collaborative and social computing}
\ccsdesc[300]{Information systems~Social networks}
\ccsdesc[100]{Applied computing~Law, social and behavioral sciences}

\keywords{YouTube; comments; hate speech; discriminatory bias}

\maketitle

\section{Introduction}
\label{sec:intro}

A recent wave of right-wing activity, including far-right and alt-right extremism, seems to be in course of progress in developed countries (especially in the United States of America~\cite{bbc} and in Western Europe~\cite{the-guardian, italy}), but also in developing countries, including Brazil~\cite{the-guardian-brazil}.
According to the Jewish non-governmental organization Anti-Defamation League (ADL), ``Internet has provided the far-right fringe with formerly inconceivable opportunities'', making it possible for extremists to reach a much larger audience than ever before and easily portray themselves as legitimate~\cite{adl}.
Analyzing how this kind of content is related to the reactions that it produces is of utmost importance to understand its peculiarities and tendencies.

YouTube, the major online video sharing website, is one of the virtual services that host a high variety of right-wing voices~\cite{vice, nyt}.
Since YouTube makes it possible for users to not only watch videos, but also to react to them through comments, it is interesting to observe how these comments are related to the content of the videos published in the platform.
It is also valuable to investigate whether behaviors connected to hate, violence and discriminatory bias come into sight in right-wing videos.
This becomes even more relevant if we consider the findings of a 2018 newspaper investigation~\cite{wallstreet} which shows that YouTube's recommendations often lead users to channels that feature highly partisan viewpoints -- even for users that have not shown interest in such content.

In this study, we analyze the content of videos published in a set of right-wing YouTube channels and observe the relationship between them and the comments that they receive from their audience using a three-layered approach in which we analyze (a) lexicon, (b) topics and (c) implicit biases present in the texts.
We also use the same approach to compare right-wing channels with a set of baseline channels in order to identify characteristics that differentiate or associate these two groups.

\paragraph{Research questions}

Our main goal is to investigate the presence of hateful content and discriminatory bias in a set of right-wing channels through the analysis of the captions of their videos and the comments posted in response to them, and to compare these captions and comments with those of a group of baseline channels.
Our initial research questions are the following:
\begin{itemize}
\item[RQ-1:] is the presence of hateful vocabulary, violent content and discriminatory biases more, less or equally accentuated in right-wing channels?
\item[RQ-2:] are, in general, commentators more, less or equally exacerbated than video hosts in an effort to express hate and discrimination?
\end{itemize}


One of the side contributions of this paper is the proposal of a three-layered method that can be used to evaluate the presence of hate speech and discriminatory bias not only on YouTube videos and comments, but in any kind of text instead.
Our method, which uses only open source tools, is an aggregation of three already established procedures that, in our view, complement each other and favor a multi-directional analysis when combined together.

This article is structured as follows:
in the next section, we describe the process of acquisition and preparation of the dataset used in our investigations; then, in Section~\ref{sec:analyses}, we detail our three analyses and present the results found; later, in Section~\ref{sec:related}, we present previous works related to the analysis of hate, violence and bias in YouTube and in online social networks in general; finally, we conclude this paper in Section~\ref{sec:conclusions} by summarizing its outcomes and by pointing out some possible future works.

\section{Data acquisition and preparation}
\label{sec:methodology}

\subsection{Dataset}

To select the YouTube channels to be analyzed, we used the website InfoWars\footnote{\url{https://www.infowars.com/}} as a seed.
InfoWars is known as a right-wing news website founded by Alex Jones, a radio host based in the United States of America.
The InfoWars website links to Alex Jones' YouTube channel, which had more than 2 million subscribers as of October 2017.
As stated in a The Guardian's article~\cite{guardian-alexjones}, ``The Alex Jones Channel, the broadcasting arm of the far-right conspiracy website InfoWars, was one of the most recommended channels in the database of videos'' used in a study which showed that YouTube's recommendation algorithm was not neutral during the presidential election of 2016 in the United States of America~\cite{guardian-1, guardian-2}.
At the moment of our data collection, Alex Jones expressed support to 12 other channels in his public YouTube profile.
We visited these channels and confirmed that, according to our understanding, all of them published mainly right-wing content.

Alex Jones' channel and these other 12 channels supported by him were then collected using the YouTube Data API\footnote{\url{https://developers.google.com/youtube/v3/}} from September 28 to October 12 2017.
From all videos posted in these channels (limited to around 500 videos per channel due to API limits), we collected (a) the~\textit{video captions} (written versions of the speech in the videos, manually created by the video hosts or automatically generated by YouTube's speech-to-text engine), representing the content of the videos themselves; and (b) the~\textit{comments} (including replies to comments)
posted to the videos.
The total number of videos collected from these channels is 3,731 and the total number of comments collected from them is 5,071,728.

In order to build a baseline set of channels to compare the results of the analyses performed in these right-wing channels with a more general behavior in YouTube videos, we collected the same information (captions and comments)
from videos posted in the ten most popular channels (in terms of number of subscribers in November 7 2017) of the category "news and politics" according to the analytics tracking site Social Blade\footnote{\url{https://socialblade.com/}}.
To be part of our baseline dataset, the content of these channels needed to be mainly in English language and non hard-coded captions needed to be available for the most part of the videos.
The total number of videos collected from the baseline channels is 3,942 and the total number of comments collected from them is 12,519,590. It is important to notice that this selection of baseline channels does not intend to represent, by any means, a ``neutral'' users' behavior (if it even exists at all).
Table~\ref{tab:videos} shows statistics regarding all collected channels.


\begin{table}[ht]
\small
\centering
\caption{Statistics regarding all collected channels.
}
 \begin{tabular}{l|r|r|r}
 \toprule
 \multirow{2}{*}{\textbf{Right-wing channels}} & \multirow{2}{*}{\textbf{Subscribers}} & \textbf{Videos} & \textbf{Comments}\\
 & & \textbf{collected} & \textbf{collected}\\
 \midrule
 The Alex Jones Channel 	& 2,157,464 		& 564 & 955,705\\
 Mark Dice 			& 1,125,052 	& 204 & 2,025,513\\
 Paul Joseph Watson 		& 1,043,236 	& 230 & 1,747,497\\
 THElNFOWARRlOR 		& 177,736 	& 467 &112,060\\
 Millennial Millie 		& 79,818 	& 359 &167,569\\
 Resistance News 		& 36,820 	& 112 & 	40,829\\
 Owen Shroyer 			& 36,125 	& 157 & 	8,000\\
 David Knight InfoWars 		& 30,940 	& 508 &1,786\\
 PlanetInfoWarsHD 		& 22,674 	& 206 &4,903\\
 Real News with David Knight 	& 12,042 	& 208 &3,902\\
 Infowars Live 			& 9,974 	& 8 &216\\
 War Room 			& 7,387 	& 188 &2,036\\
 Jon Bowne Reports 		& 5,684 	& 520 & 	1,712\\
 \midrule
 \textbf{Total} & 4,744,925 & 3,731 & 5,071,728\\
 \midrule
 \midrule
 \multirow{2}{*}{\textbf{Baseline channels}} & \multirow{2}{*}{\textbf{Subscribers}} & \textbf{Videos} & \textbf{Comments}\\
 & & \textbf{collected} & \textbf{collected}\\
 \midrule
 YouTube Spotlight &25,594,238 &262 &734,591 \\
 The Young Turks &3,479,018 &540 &1,652,818\\
 Barcroft TV &3,459,016 &427 &1,279,400\\
 Vox &3,103,138 &448 &1,389,170\\
 DramaAlert &3,081,568 &470 &4,904,941\\
 VICE News &2,476,558 &451 &897,056\\
 YouTube Spotlight UK &2,307,818 &75 &23,280\\
 TomoNews US &1,928,700 &543 &338,501\\
 SourceFed &1,713,646 &501 &838,431\\
 Anonymous Official &1,700,812 &225 &461,402\\
 \midrule
 \textbf{Total} & 23,275,686 & 3,942 & 12,519,590\\
 \bottomrule
 \end{tabular}
 \label{tab:videos}
\end{table}

\subsection{Textual preprocessing}
\label{sec:preprocessing}

First, HTML tags and URLs were removed from both video captions and users' comments. Also, we used~\texttt{langid.py}\footnote{\url{https://github.com/saffsd/langid.py}}~\cite{lui2012langid}, a language identification tool, to filter only video captions and comments with a probability $\geq$0.8 of being in English.
This filtering resulted in the 3,278 videos and 4,348,986 comments from right-wing channels and in the 3,581 videos and 9,522,597 comments from baseline channels used in our investigations.
Then, for each video we created two documents, each one originating from one of the two sources (\emph{caption} and
\emph{comments}).

When additional preprocessing stages were required for an analysis, we mention them in the subsection corresponding to the specific methodology of that analysis, in Section~\ref{sec:analyses}.

\section{Analyses and results}
\label{sec:analyses}

\begin{figure*}[ht!]
\centering
\includegraphics[width=1.0\textwidth]{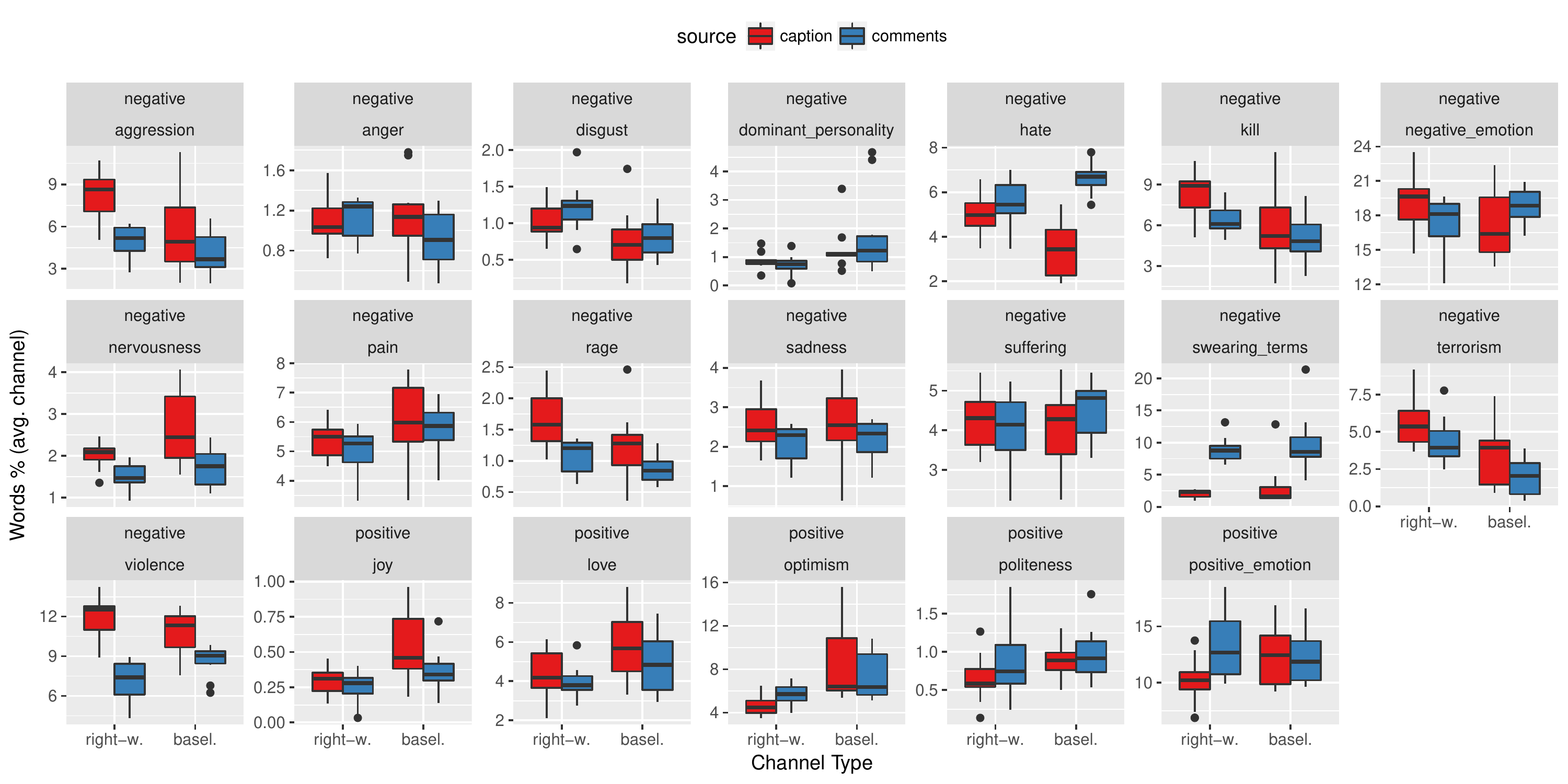}
\caption{Normalized percentage of words in each semantic field represented by an Empath category. The bottom and top of the box are always the first and third quartiles, the band inside the box is the median, the whiskers represents the minimum and maximum values, and the dots are outliers.} 
\label{fig:lexcatB-bychannel-boxplot}
\end{figure*}



We use a three-layered approach to investigate the problem of hate, violence and discriminatory bias in our set of right-wing videos and to address the research questions formulated in Section~\ref{sec:intro}.
Our three analyses, through which we evaluate (a) lexicon, (b) topics and (c) implicit biases, are the following:

\begin{itemize}
\item \textbf{lexical analysis}: we compared the semantic fields of the words in the captions with the semantic fields of the words in the comments, focusing on semantic fields related to hate, violence and discrimination. We did the same to compare right-wing channels to baseline channels;
\item \textbf{topic analysis}: we contrasted the topics addressed in the captions with the ones addressed in the comments. Again, we did the same to contrast right-wing channels to baseline channels;
\item \textbf{implicit bias analysis}: we analyzed implicit biases based on vector spaces in which words that share common contexts are located in close proximity to one another. Through this method, we compared biases between captions and comments, and once again between right-wing and baseline channels.
\end{itemize}

\subsection{Lexical analysis}\label{subsec:lexical}

Lexical analysis, that is, the investigation of the vocabulary, reveals how society perceives reality and indicates the main concerns and interests of particular communities of speakers~\cite{cunha14_ht}.
According to lexicological theories, vocabulary is the translation of social realities and thus it is natural to study it as a means to comprehend characteristics of groups that employ certain words in their discourse~\cite{matore53,cambraia2013lexicologia}.
Several different ways of analyzing vocabulary are possible.
In this study, we model each channel based on the semantic fields (i.e. groups of semantically related items) of the words used in its videos and in the comments that it received.

\subsubsection{Methodology}
\label{sec:lexcat_methodology}

In addition to the preprocessing tasks mentioned in Section~\ref{sec:preprocessing}, lemmatization was applied by employing the WordNet Lemmatizer function provided by the Natural Language Toolkit~\cite{bird2009natural} and using~\textit{verb} as the part-of-speech argument for the lemmatization method.
For this analysis, lemmatization was necessary in order to group together the inflected forms of the words, so they could be analyzed as single items based on their dictionary forms (\textit{lemmas}).
In this way, words like~\textit{cat} and~\textit{cats} were grouped together under the same lemma (in this case,~\textit{cat}).

Then, each word was classified according to categories that represent different semantic fields, such as diverse topics and emotions, provided by Empath~\cite{fast2016empath}, ``a tool for analyzing text across lexical categories''~\footnote{\url{https://github.com/Ejhfast/empath-client}}.
From the 194 total Empath categories, we selected the following (a) 15 categories related to hate, violence, discrimination and negative feelings, and (b) 5 categories related to positive matters in general:
\begin{itemize}
	\item \textbf{negative:}~\textit{aggression, anger, disgust, dominant personality, hate, kill, negative emotion, nervousness, pain, rage, sadness, suffering, swearing terms, terrorism, violence}
	\item \textbf{positive:}~\textit{joy, love, optimist, politeness, positive emotion}
\end{itemize}
For a given video video $v$, we calculated the word count for each one of these selected categories as
\begin{equation}\label{eq:E}
\vec{E}_{v, \text{source}} = (e_1, e_2, \dots, e_{19}, e_{20})\,,
\end{equation}
where $e_i$ is the number of words from category $i$, and~\textit{source} is either $caption$ or $comments$, resulting in two vectors for each video.
Since the videos vary in terms of size and number of comments, we also created normalized vectors, defined for a video $v$ as 
\begin{equation}\label{eq:EN}
\vec{EN}_{v, \text{source}} = \frac{\vec{E}_{v, \text{source}}}{\sum_{i=1}^{20} e_i} = \left(\frac{e_1}{\sum e_i}, \frac{e_2}{\sum e_i}, \dots, \frac{e_{19}}{\sum e_i}, \frac{e_{20}}{\sum e_i} \right)\,,
\end{equation}
which contain the normalized fraction of words presented in each Empath category. Again, for each video we have two normalized vectors: one for its captions and another one for its comments.

In order to have an unique vector representing an entire channel (instead of a single video only), we defined an average vector that aggregates all videos of that particular channel. For a given channel $c$, we define
\begin{align}\label{eq:EC}
\vec{EC}_{c, \text{source}} = (ec_1, ec_2, \dots, ec_{19}, ec_{20}) \nonumber \\ 
ec_i = \frac{\sum_{v \in V_c} \left( \vec{EN}_{v, \text{source}}[i] \right) }{|V_c|}\,,
\end{align}
where $V_c$ is the set of all videos of a channel $c$. In words, the vector $\vec{EC}_{c}$ contains the average fraction of each Empath category present in the caption or in the comments of the videos in channel $c$.

Finally, we defined a metric that measures the similarity between content and comments of a video. This metric measures the cosine similarity~\cite{Singhal01moderninformation} between the two vectors of a particular video $v$ and is defined as 
\begin{equation}\label{eq:S}
S_v = \cos \left( \vec{EN}_{v, caption},\ \vec{EN}_{v, comments} \right)\,.
\end{equation}
Since our vectors do not hold negative values, the cosine similarity between them varies from 0 (totally different) to 1 (identical).

\subsubsection{Results}
\paragraph{Comparing semantic fields between channel types and between sources} First, we analyze the semantic fields present in each channel type (right-wing vs. baseline) and those arising from each source (caption vs. comments). As explained above, we computed two normalized vectors of percentage for each video, and then calculated the average value for each channel ($\vec{EC}_{caption}$, $\vec{EC}_{comments}$).

Figure~\ref{fig:lexcatB-bychannel-boxplot} depicts the normalized percentage of words in each semantic field represented by an Empath category. 
We observe a clear and consistent dominance of some negative categories, including~\textit{nervousness},~\textit{rage} and~\textit{violence}, among captions (if compared to comments). On the other hand, comments contain predominantly more~\textit{swearing terms}. Interestingly, for the category~\textit{hate}, while there is no significant difference for right-wing channels, for the baseline channels there is a considerable difference between captions and comments: median of 3.5\% vs. 6.8\%, respectively, thus reporting a percentage of~\textit{hate} for baseline comments even greater than for right-wing comments.

Comparing channel types, we observe that right-wing channels have higher fractions of words from other negative categories, such as~\textit{disgust},~\textit{kill} and~\textit{terrorism}, while baseline channels present higher fractions of positive categories such as~\textit{joy} and~\textit{optimism} (although also presenting higher fraction for the category~\textit{pain}). It is also worth noting categories that show no statistical difference between channel types, like~\textit{disgust} and~\textit{swearing terms}. 
Another interesting result regards the category~\textit{positive emotion}: although there is no statistical difference between baseline's captions and comments, the same is not true for right-wing channels, for which there are more words of this category in comments than in captions.

\begin{figure}[ht!]
\centering
\includegraphics[width=0.5\textwidth]{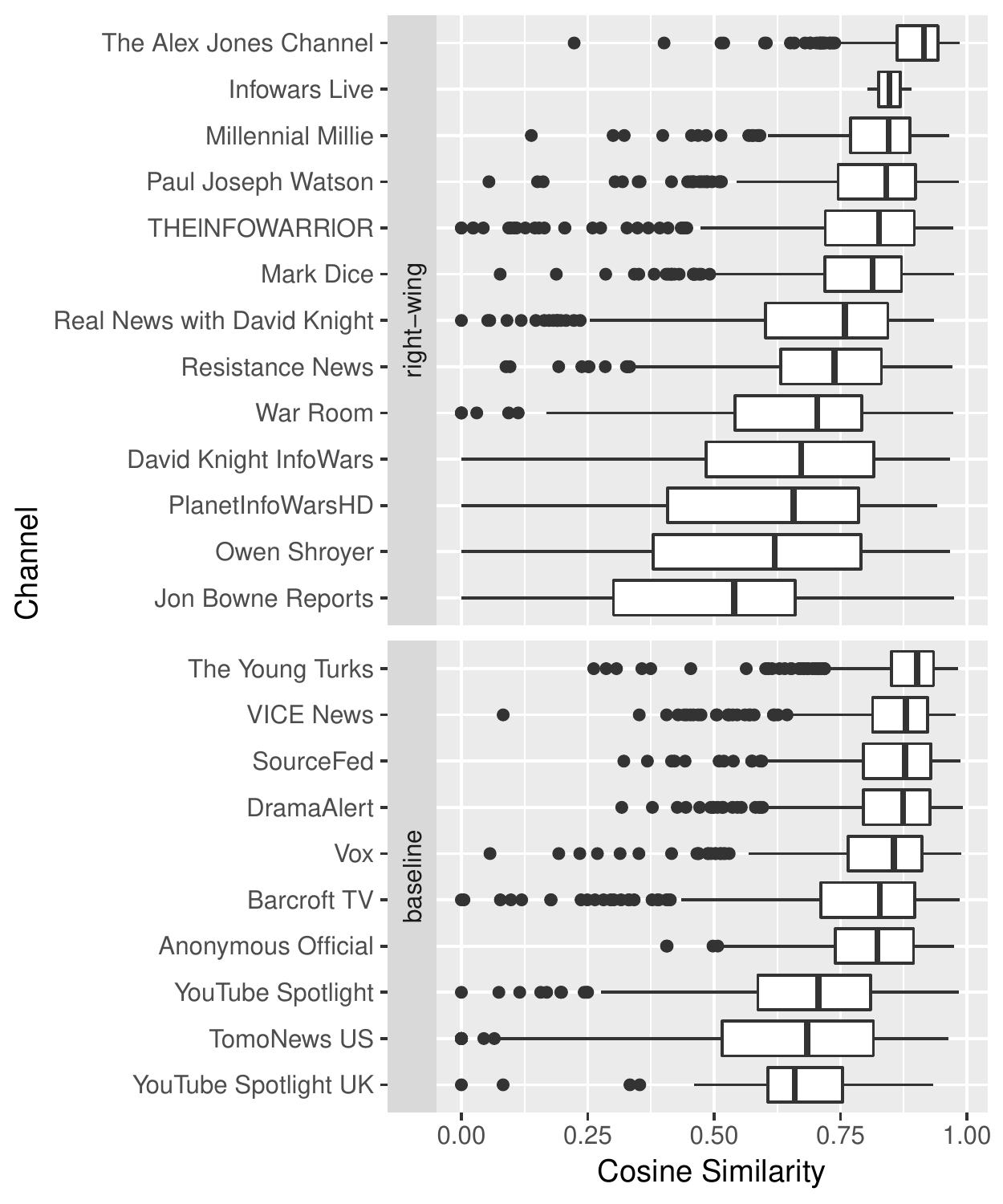}
\caption{Distribution of the similarities between caption and comments in each channel's videos, according to our lexical analysis. Values close to 0 indicate no correlation and values close to 1 report maximum correlation.}
\label{fig:lexcatB-similarity}
\vspace{-0.4cm}
\end{figure}

\paragraph{Similarity between caption and comments} Now, we compare the similarity between the semantic fields present in the caption and in the comments of a given video $v$ by calculating the previously defined metric $S_v$. Figure~\ref{fig:lexcatB-similarity} depicts the boxplot distribution of this similarity in each channel's videos. 

We notice a high variation among the similarity values in videos of a same channel: while in some videos the occurring semantic fields in the host's discourse (represented by the caption) and in the audience's speech (represented by the comments) are very similar, in others the similarity can be close to zero.

This similarity also varies among channels. For instance, while ``The Alex Jones Channel'' holds a median similarity of 0.9, the median similarity in videos at ``Jon Bowne Reports'' is as low as 0.5. Interestingly, the variance of the distributions for the baseline channels is lower than the one for right-wing channels, meaning that the former generally have more consistent levels of similarity between caption and comments. It is important to notice that
it seems to exist a correlation between a channel's popularity and the similarity between the semantic fields occurring in the captions of its videos and the ones occurring in the comments of its videos: more popular channels (according to Table~\ref{tab:videos}) generally present higher values of similarity. This could be an explanation for the higher and more consistent values of similarity among baseline channels, since all of them had at least 1,700,000 subscribers at the moment of our data collection.

\begin{figure*}[ht!]
\centering
\includegraphics[width=0.9\textwidth]{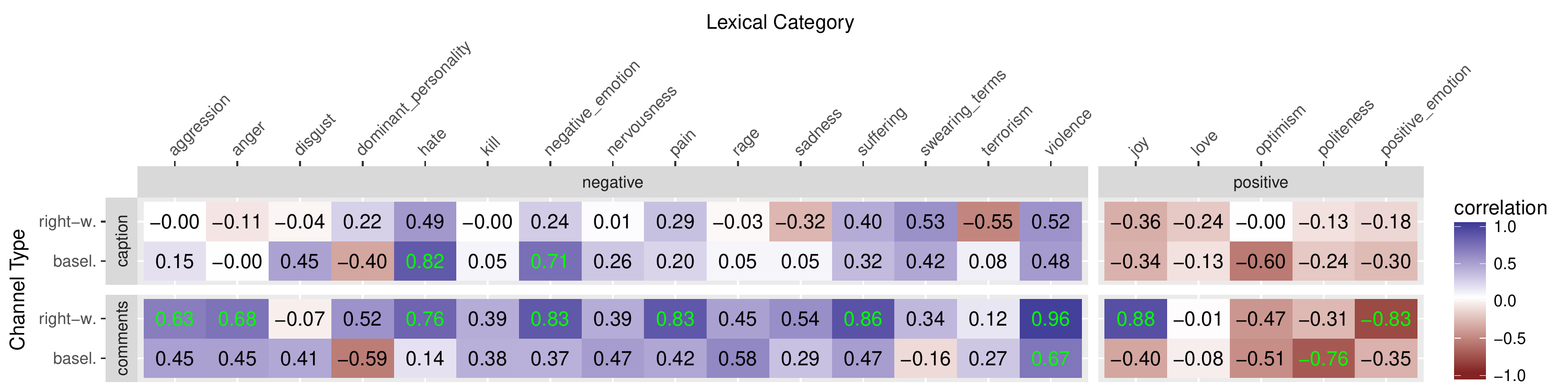}
\caption{Correlations between normalized frequencies of words in each Empath category and the average cosine similarity between the vocabulary of all captions of a channel and the vocabulary of all comments published in this channel's videos. Highlighted values indicate correlations with $p$-value<0.05 .} 
\label{fig:lexcatB-bychannel-cortable}
\end{figure*}

\paragraph{Correlation between channel's similarity and semantic fields}
Finally, we focus on identifying characteristics that could explain the levels of lexical similarity between the host and the commentators. To do that, we measured the correlation between the average similarity and the average fractions of Empath categories (that is, the dimensions of $\vec{EC}$) using the Pearson correlation coefficient. We measured the average fraction of both captions and comments, also aggregating the channels by type (right-wing and baseline).
We present the correlation values in Figure~\ref{fig:lexcatB-bychannel-cortable}, highlighting the significant correlation values (with $p$-value\textless 0.05). 

Regarding the captions, we observe no significant correlation for the right-wing channels, and a significant positive correlation for the categories~\textit{hate} and~\textit{negative emotion} for the baseline channels. These results imply that baseline channels with higher fraction of words related to hate and negative emotions also have a higher degree of similarity between caption and comments.

Now considering the comments, we observe a significant positive correlation for several categories in the right-wing channels, such as~\textit{agression},~\textit{hate} and~\textit{violence}. There is also a negative correlation for the~\textit{positive emotion} category, meaning that channels with less words related to positive emotions in their comments hold a higher similarity. By looking at the baseline channels, we only detect a significant positive correlation for~\textit{violence}, resembling right-wing channels, but with lower intensity. There is, though, a significant negative correlation for~\textit{politeness}, implying that channels with a lower fraction of these words in their comments hold a higher similarity.

\subsection{Topic analysis}

In subsection~\ref{subsec:lexical}, we address a lexical analysis of our textual corpora by studying the semantic fields of the words employed in the captions and in the comments of the videos posted in right-wing and baseline channels. Now, we employ latent Dirichlet allocation (LDA)~\cite{blei2003latent}, a way of automatically discovering topics contained in textual datasets, to investigate latent topics present in these videos' captions and comments. 




\subsubsection{Methodology}

For this analysis, beyond the preprocessing steps mentioned in Section~\ref{sec:preprocessing}, we also removed punctuation, multiple white spaces and stop words\footnote{Using the list of stop words suggested by the Python library~\texttt{gensim} in~\url{https://github.com/RaRe-Technologies/gensim/blob/develop/gensim/parsing/preprocessing.py}}. We lowercased and tokenized the whole corpus as well.


We ran the LDA algorithm using the implementation provided by~\texttt{gensim}~\cite{rehurek_lrec}, ``a Python library for topic modeling, document indexing and similarity retrieval with large corpora''~\footnote{\url{https://radimrehurek.com/gensim/}}. Due to limitations of~\texttt{gensim}'s parallel LDA implementation, we randomly selected a maximum of 2,000 tokens for each document. We chose the parameters $\alpha = \beta = 1.0/num\_topics\ prior$ and $k = 300$. The parameter $k$ indicates the number of topics to be returned by the algorithm, so our LDA model returned 300 topics, each one containing words ordered by importance in that topic. 
With a trained LDA model, we then assigned a topic to each document by generating a topic distribution for both the video's caption and comments, and then selected the most likely topic as the representative of this document.



\subsubsection{Results}

Table~\ref{tab:lda} shows a partial output of our LDA model by displaying the top 2 topics for each document and the top ranked 20 words produced by the LDA. As frequently, the words concerning each topic inferred by LDA are not strongly cohesive among each other, and are not very conclusive. Another problem is that a topic word can have multiple connotations, so that its interpretation is ambiguous. In any case, we discuss possible interpretations of the topics through a qualitative observation of the word lists.

\begin{table}[!ht]
\small
\centering
\caption{Top 2 topics for each document. Inside each topic, 20 words are presented in order of importance according to the LDA output.}
 \begin{tabular}{c|c|c}
 \toprule 
  \textbf{Document}& \specialcell[c]{\textbf{Topic}\\ \textbf{rank}} & \textbf{Topic words} \\
 \midrule
 \multirow{6}{*}{\specialcell[c]{\textbf{Right-wing}\\ \textbf{captions}}} & 1 &\specialcell[c]{vaccine, vaccines, vox, cenk, ukraine, millie,\\ flight, nato, bike, morgan, infrastructure,\\fluoride, keem, ukrainian, labour, israeli,\\torture, jeremy, awards, bombing} \\
 && \\
  & 2 & \specialcell[c]{abortion, solar, assange, kelly, wikileaks,\\petition, vox, beck, sheriff, jinx, react,\\petitions, owen, syrian, nfl, arpaio,\\rushmore, document, pregnancy, oath} \\
 \midrule
  \multirow{6}{*}{\specialcell[c]{\textbf{Right-wing}\\ \textbf{comments}}} &1 &  \specialcell[c]{quot, rays, speaker, ebola, gamma,\\palestinians, cruz, ksi, radiation, virus,\\ray, maher, candace, ted, palestinian,\\memes, ukraine, keem, irish, dnc} \\
 && \\
  & 2 &\specialcell[c]{millie, quot, owen, korean, gangs, ricegum,\\ manifesto, rice, drone, rainbow, depression,\\ discrimination, flu, speaker, feminists, jay,\\ radiation, professor, dodger, cook'} \\ 
 \midrule
  \multirow{6}{*}{\specialcell[c]{\textbf{Baseline}\\ \textbf{captions}}} & 1 &\specialcell[c]{gt, quot, whale, n, pluto, puerto, horizons,\\ loopholes, irish, rico, playlist,\\nasa, sheriff, axis, maryanne,\\megyn, swamp, faze, vox, surface} \\
 && \\
  & 2 &\specialcell[c]{gt, commentary, hurricane, papa, sarry,\\kevin, quot, ali, fifa, n, hammer,\\cenk, wolf, donors, symbols, shark,\\keem, trudeau, starbucks, warren} \\
 \midrule
  \multirow{6}{*}{\specialcell[c]{\textbf{Baseline}\\ \textbf{comments}}} & 1 & \specialcell[c]{keem, rice, ricegum, leafy, dramaalert,\\scarce, faze, squad, lizard, pewdiepie,\\rap, rain, idubbbz, keems, michelle,\\diss, bleach, subbed, quantum, ty} \\
 && \\
  & 2 &\specialcell[c]{dan, cenk, phil, ana, bees,\\keem, millie, bee, leafy, quot,\\minecraft, mars, generic, turks, roger,\\antifa, ava, todd, flight, feminists} \\
 \bottomrule
 \end{tabular}
 \label{tab:lda}
\end{table}

Among the top ranked topics for the right-wing captions, we observe a relevant frequency of words related to war and terrorism, including~\textit{nato},~\textit{torture} and~\textit{bombing}, and a relevant frequency of words related to espionage and information war, like~\textit{assange},~\textit{wikileaks}, possibly~\textit{document} and \textit{morgan} (due to the actor Morgan Freeman's popular video in which he accuses Russia of attacking United States' democracy during its 2016 elections\footnote{\url{http://bbc.in/2BQljyP}}).

Regarding the top ranked topics for the right-wing comments, it is possible to recognize many words probably related to biological and chemical warfare, such as~\textit{rays},~\textit{ebola},~\textit{gamma},~\textit{radiation} and~\textit{virus}. It is also interesting to observe the presence of the word~\textit{palestinian} in the highest ranked topic: it might indicate that commentators are responding to the word~\textit{israeli}, present in the top ranked topic of the captions.





As expected, the words in the  top ranked topics of the baseline channels seem to cover a wider range of subjects.
The terms in the top ranked topics of the baseline captions include words regarding celebrities, TV shows and general news, while the ones in the baseline comments are very much related to Internet celebrities such as~\textit{RiceGum} and~\textit{PewDiePie}, and computer games, like~\textit{Minecraft}. In the second highest ranked topic, however, we also observe a small political interest through the presence of the words~\textit{antifa} and~\textit{feminists}.






We observe that, in general, topics in the captions and comments of right-wing channels are more specific than those of our baseline channels. This was somewhat expected, since our baseline dataset is composed of channels about varied topics and general interests.

\subsection{Implicit bias analysis}

After investigating vocabulary and topics, we now move up one more level of analysis and observe implicit discriminatory biases that can be retrieved from our dataset of video captions and comments.

The~\textit{Implicit Association Test} (IAT) was introduced by~\citet{greenwald1998measuring} to study unconscious, subtle and often unintended biases in individuals. Its core idea is to measure the strength of associations between two target concepts (e.g.~\textit{flowers} and~\textit{insects}) and two attributes (e.g.~\textit{pleasant} and~\textit{unpleasant}) based on the reaction time needed to match (a) items that correspond to the target concepts to (b) items that correspond to the attributes (in this case,~\textit{flowers + pleasant},~\textit{insects + pleasant},~\textit{flowers + unpleasant},~\textit{insects + unpleasant}). The authors found that individuals' performance was more satisfactory when they needed to match implicit associated categories, such as~\textit{flowers + pleasant} and~\textit{insects + unpleasant}.



\citet{caliskan2017semantics} propose applying the IAT method to analyze implicit biases based on vector spaces in which words that share common contexts are located in close proximity to one another, generated by a technique called~\textit{word embedding}. By replicating a wide spectrum of biases previously assessed by implicit association tests, they show that cosine similarity between words in a vector space generated by word embeddings is also able to capture implicit biases. The authors named this technique~\textit{Word Embedding Association Test} (WEAT).



\subsubsection{Methodology}\label{weat-metodology}


We created three WEATs focused on harmful biases towards the following minorities and/or groups likely to suffer discrimination in North America and Western Europe: immigrants, LGBT people and Muslims. The words that compose each class and attribute in our tests are shown in Table~\ref{tab:WEAT}. According to~\citet{caliskan2017semantics}, the two classes to be evaluated must contain the same number of words, but the sizes of the sets of attributes can be different. Words from ``Class 1'' are related to discriminated groups, while words from ``Class 2'' concern dominant groups; attributes from ``Attributes 1'' are negative elements and attributes from ``Attributes 2'' are positive elements.

\begin{table}[!ht]
\small
\centering
\caption{Words that compose each class and set of attributes in our Word Embedding Association Tests (WEATs).}
 \begin{tabular}{c|c|c|c}
 \toprule 
 & \textbf{Immigrants} & \textbf{Muslims} & \textbf{LGBT people}\\
 \midrule
 \specialcell[c]{\textbf{Class 1}\\(discriminated)} & \specialcell[c]{immigrant,\\migrant} & \specialcell[c]{islamism,\\muhammed,\\muslim, quran} & \specialcell[c]{bisexual, gay,\\homosexual,\\lesbian}\\ 
 \midrule
 \specialcell[c]{\textbf{Class 2}\\(dominant)} & \specialcell[c]{citizen,\\native} & \specialcell[c]{bible, christian,\\christianity,\\jesus} & \specialcell[c]{het, hetero,\\heterosexual,\\straight}\\
 \midrule
 \specialcell[c]{\textbf{Attributes 1}\\(negative)} & \specialcell[c]{bad, burden,\\pirate, plague,\\taker, thief} & \specialcell[c]{assassin, attack,\\bomb, death,\\murder, radical,\\terrorist} & \specialcell[c]{immoral,\\outrageous,\\promiscuous,\\revolting, sinner}\\ 
 \midrule
 \specialcell[c]{\textbf{Attributes 2}\\(positive)} & \specialcell[c]{good, honest,\\maker, rightful} & \specialcell[c]{compassionate,\\gentle, humane,\\kind, tolerant} & \specialcell[c]{moral, natural,\\normal}\\
 \bottomrule
 \end{tabular}
 \label{tab:WEAT}
\end{table}




Then, we used a collection containing all the articles of Wikipedia's English-language edition\footnote{Downloaded in March 5 2017 and available at~\url{https://dumps.wikimedia.org/}} to pre-train a base model with 600 dimensions employing~\texttt{word2vec}\footnote{\url{https://code.google.com/archive/p/word2vec/}}\cite{mikolov1,mikolov2}.
We chose to use data from Wikipedia due to its popularity as a base model for language modeling applications using word embeddings, since it is a large dataset often considered to be a good representation of contemporary English~\cite{mesnil2013investigation,levy2014neural}. Also, due to limited access to domain-specific text corpora (in our case, captions and comments from right-wing YouTube channels), it is beneficial to initialize the models with weights and vocabulary trained in a large text corpus and then re-train the weights with the domain-specific dataset~\cite{kusner2015word,sienvcnik2015adapting,collobert2011natural}.


Once the Wikipedia base model was created, we used it as the starting point for our specific models. For each YouTube channel in our dataset, we trained two~\texttt{word2vec} models: one of them concerning the captions and the other one concerning the comments in the videos. Then, we implemented our WEATs according to the method proposed by~\citet{caliskan2017semantics}, that is, measuring (a) the association between a given word $w$ and the attributes $A_1$ and $A_2$ (Equation~\ref{eq:a}), and (b) the association between the two sets of target words belonging to the classes $C_1$ and $C_2$ and the two sets of attributes $A_1$ and $A_2$ (Equation~\ref{eq:b}), as in

\begin{equation}\label{eq:a}
s(w,A_1,A_2) = Mean_{a \epsilon A_1}(\cos(\vec{w},\vec{a})) - Mean_{b \epsilon A_2}(\cos(\vec{w},\vec{b}))
\end{equation}
and
\begin{equation}\label{eq:b}
s(C_1,C_2,A_1,A_2) = \sum_{x \epsilon C_1} s(x,A_1,A_2) - \sum_{y \epsilon C_2} s(y,A_1,A_2)\,,
\end{equation}
where $\cos(\vec{x},\vec{y})$ indicates the cosine of the angle between the vectors $\vec{x}$ and $\vec{y}$. The effect sizes of these associations are the normalized measures of how separated the two distributions of associations between classes and attributes are, and are calculated through Cohen's $d$, which, in this case, is defined as

\begin{equation}
d = \frac{Mean_{x \epsilon C_1}(s(x,A_1,A_2)) - Mean_{y \epsilon C_2} ( s(y,A_1,A_2))}{\sigma_{w \epsilon C_1 \bigcup C_2} s(w,A_1,A_2)}\,,
\end{equation}
where $\sigma$ stands for the standard deviation. The significance of the effect sizes are represented by $p$-values calculated asserting the one-sided permutation test using all the possible partitions of the two classes into two sets of equal size ($X_i,Y_i$). In this case, the $p$-value is defined as the probability that one of these possible permutations yields a test statistic value greater than the one observed by our WEAT definitions in Table~\ref{tab:WEAT}:

\begin{equation}
P_{value} = Pr(s(X_i,Y_i,A_1,A_2) > s(C_1,C_2,A_1,A_2))\,.
\end{equation}


%
%


\subsubsection{Results}\label{subsec:weat-results}

\begin{figure*}[ht!]
\centering
\includegraphics[width=0.9\textwidth]{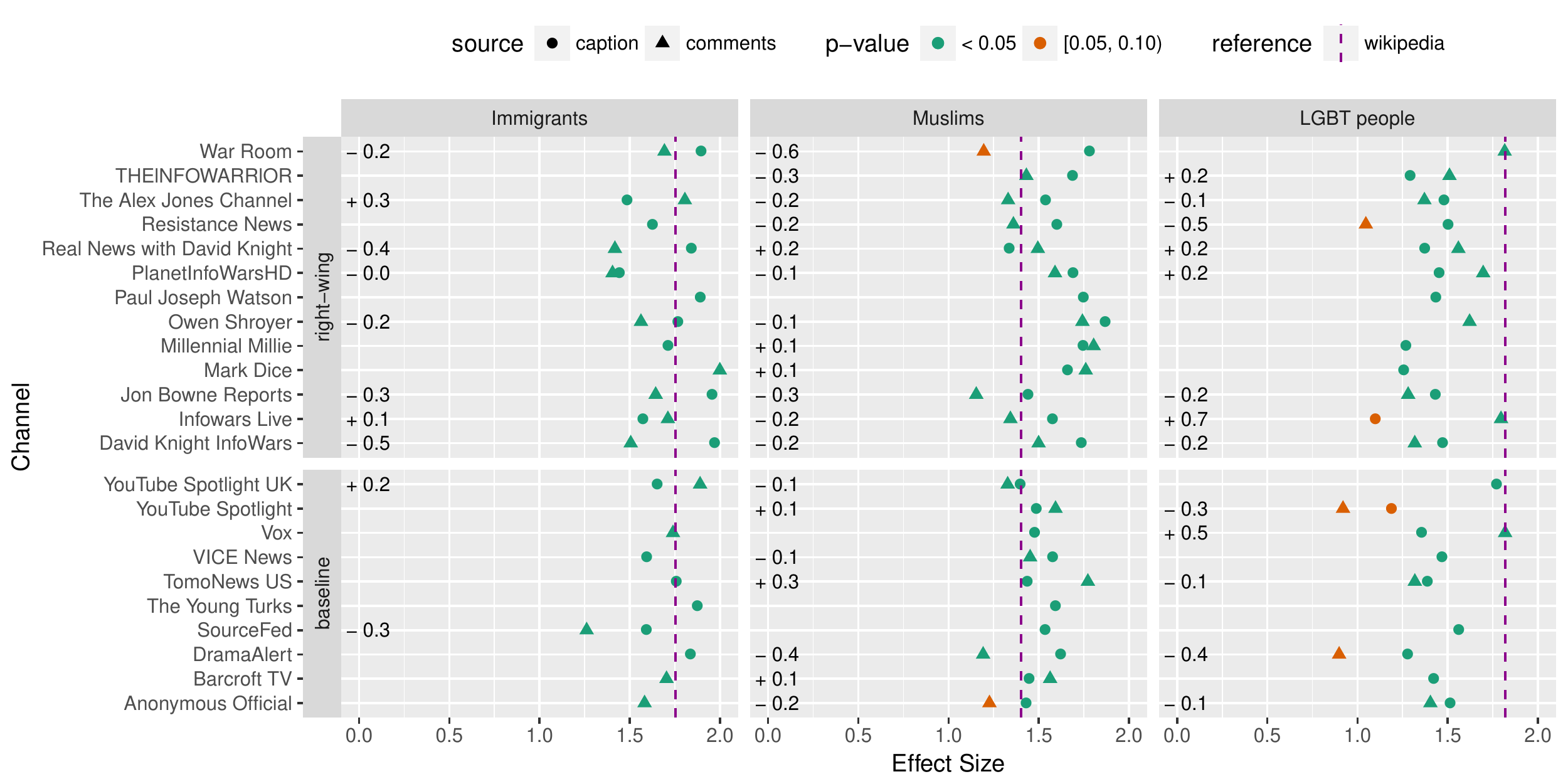}
\caption{Value of WEAT biases for the three topics analyzed. Dashed lines indicate the reference value calculated from the Wikipedia corpus. The numbers indicate the difference between biases calculated for comments and captions.} 
\label{fig:weat_points}
\end{figure*}

\begin{figure}[ht!]
\centering
\includegraphics[width=0.48\textwidth]{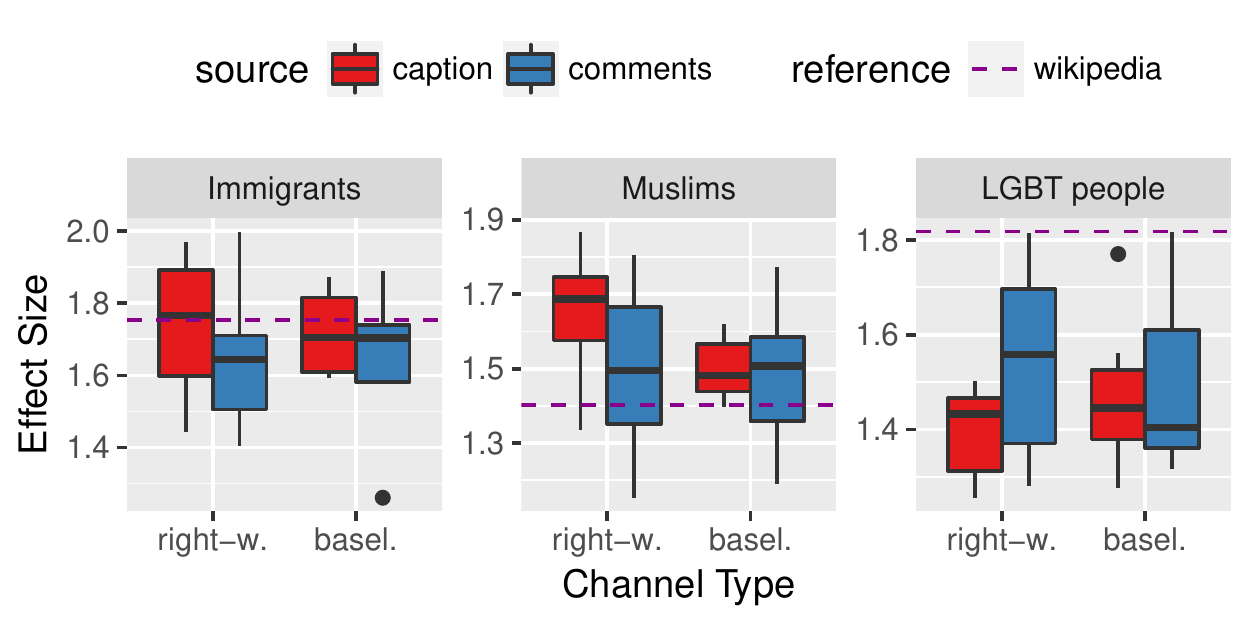}
\caption{Distribution of WEAT biases for the three topics analyzed. Dashed lines indicate the reference value calculated from the Wikipedia corpus.} 
\label{fig:weat_boxplot}
\end{figure}

We present in Figure~\ref{fig:weat_points} the values of biases of the three topics in terms of effect size (Cohen's $d$) for all the right-wing and baseline channels, both for captions and comments. In the plot, we only show the biases with $p$-value\textless $0.1$, being the ones in the range $[0.05, 0.1)$ in orange and the ones~\textless $0.05$ in green. The dashed line is a reference value indicating the bias present in the Wikipedia corpus alone. The signed numbers indicate the difference of bias between comments and captions, where a positive value represents a higher bias for comments and a negative value indicates a higher bias for caption. We also depict, in Figure~\ref{fig:weat_boxplot}, the boxplot of these values, aggregating for channel type and source, and considering only the biases with p-value\textless $0.05$.

\paragraph{Comparing channels' implicit biases with Wikipedia corpus}


First, we highlight that, according to our WEATs, the baseline Wikipedia corpus holds a relatively high bias by itself. 
This is consistent with previous studies~\cite{bullinaria2007extracting,stubbs1996text,caliskan2017semantics}, indicating that cultural biases are transmitted through written language. 

When contrasting the reference Wikipedia bias with the YouTube biases, we observe different trends depending on the topic. For instance, the bias against Muslims was almost always amplified when compared to the reference, especially for captions. On the other hand, bias against LGBT people was weakened in most of the observed channels, even in the right-wing ones.
Concerning the bias against immigrants, the values appear close
to the reference.


\paragraph{Comparing biases in captions with biases in comments} 


It is interesting to notice that, for immigrants and Muslims, captions hold higher biases than comments in 75\% of the right-wing channels, considering the statistically significant cases (Figure~\ref{fig:weat_points}). The fact that, in right-wing channels, comments hold lower bias against immigrants and Muslins when compared to captions can also be stated by looking at Figure~\ref{fig:weat_boxplot}. For LGBT people, however, comments hold higher discriminatory bias in right-wing channels.

\paragraph{Comparing right-wing and baseline biases} 


We observe that, concerning Muslims, the captions of right-wing channels present higher biases (median $ = 1.7$) than baseline channels (median $ = 1.5$). For the other topics, the differences were not very pronounced. It is also worth to mention that, as shown in Table~\ref{tab:WEAT}, the fraction of channels with statistically significant biases is much higher for right-wing channels, regardless of source (captions or comments). For this reason, for  many baseline cases, we cannot conclude that a significant difference exists, nor conclude that it does not exist.

%
%
%
%
%
%


\subsection{Multi-layered analysis}


Now, we summarize the findings of each of the three previous analyses and combine their results in order to answer the research questions proposed in Section~\ref{sec:intro}.

\paragraph{RQ-1: is the presence of hateful vocabulary, violent content and discriminatory biases more, less or equally accentuated in right-wing channels?} Our lexical analysis shows that right-wing channels, when compared with baseline channels, incorporate higher percentages of words conveying semantic fields like~\textit{aggression},~\textit{kill},~\textit{rage} and~\textit{violence}, while baseline channels hold a higher percentage of positive semantic fields such as~\textit{joy} and~\textit{optimism}.
Even though the most frequent LDA topics do not show high evidences of hate, they did report that right-wing channels debates are more related to subjects like war and terrorism, which might corroborate the lexical analysis. 
Also, the implicit bias analysis shows that, independently of channel type (right-wing or baseline), the YouTube community seems to amplify a discriminatory bias against Muslims, depicted as assassins, radicals and terrorists, and weaken the association of LGBT people as immoral, promiscuous and sinners when compared to the Wikipedia reference. 

Although the lexical and topic analysis show evidences of negative feelings, they are unable to indicate towards whom these feelings are addressed.
The implicit bias analysis shows no differences between right-wing and baseline captions regarding immigrants and LGBT people, but it does show against Muslims. 
We might conclude, then, that hateful vocabulary and violent content seems to be more accentuated in right-wing channels than in our set of baseline channels, and also that a discriminatory bias against Muslims is more present in right-wing videos. 

\paragraph{RQ-2: are, in general, commentators more, less or equally exacerbated than video hosts in an effort to express hate and discrimination?} 
The lexical analysis reports that comments generally have more words from the semantic fields~\textit{disgust},~\textit{hate} and~\textit{swearing terms}, and captions express more~\textit{aggression},~\textit{rage} and~\textit{violence}. Regarding biases against immigrants and Muslims, in 75\% of the  right-wing channels the comments show less bias than the captions. On the other hand, although the implicit bias against LGBT people in YouTube is generally lower than in the Wikipedia reference, it is greater on right-wing comments than in right-wing captions.

Our conclusion is that, in general, YouTube commentators are more exacerbated than video hosts in the context of hate and discrimination, even though several exceptions may apply.

\section{Related work}
\label{sec:related}

\paragraph{On hate, violence and bias on the Web}

The analysis of hate, violence and discriminatory bias in online social networks is gaining a lot of attention in the field of social computing as platforms such as Facebook, Instagram and Twitter, to name a few, connect more and more users at a global level -- being one of the topics covered by what has been called~\emph{computational social science}~\cite{lazer2009life}.

The identification of hateful messages in online services is still an open question.
Schmidt and Wiegand~\cite{schmidt2017survey} show that the manual inspection of hateful content in a social media service is not feasible, and present a survey describing key areas on natural language processing that have been explored to automatically recognize hateful content. 
Ribeiro et al.~\cite{ribeiro2017like} propose a different approach, focusing on a user-centric view of hate speech and characterizing hateful Twitter users instead of hateful messages.
The authors show that these users tend to be more negative, more profane and, counter-intuitively, use less words associated with topics such as hate, terrorism, violence and anger.

Hate and violence in the video sharing website YouTube is also increasingly receiving scholarly attention.
Sureka et al.~\cite{sureka2010mining} propose a solution based on data mining and social network analysis to discover hate videos, users and virtual hidden communities on YouTube, while Agarwal and Sureka~\cite{agarwal2014focused} present a focused-crawler based approach for mining hate and extremism in this social platform.

Case studies are also useful for the purpose of elucidating the dynamics and the strength of online activity related to hate, violence and discriminatory bias.
For instance, Chatzakou et al.~\cite{DBLP:journals/corr/ChatzakouKBCSV17a} investigate the behavior of users involved in the~\emph{Gamergate controversy}, a harassment campaign against women in the video game industry that lead to many incidents of cyberbullying and cyberaggression.
The authors compare the behavior of Twitter users considered~\emph{gamergaters} with the behavior of baseline users, finding that gamergaters ``post tweets with negative sentiment, less joy, and more hate than random users''.
On another vein, Savage and Monroy-Hern{\'a}ndez~\cite{savage2015participatory} analyze a militia uprising unfolded on social media in the context of the Mexican War on Drugs, illustrating its ``online mobilization strategies, and how its audience takes part in defining the narrative of this armed conflict''.

\paragraph{On comment behavior on the Web}

The behavior of commentators in websites and in online social media services is also a growing research topic in social computing.
Through the analysis of interviews with frequent Internet commentators, French~\cite{french} shows that the reasons for users to comment on websites are many and varied.
Stroud, Van Duyn and Peacock~\cite{stroud2} indicate that social media is the most prevalent place for Internet users to comment and read comments. They add that most commentators and comment readers ``agree that allowing anonymity in comment
sections allows participants to express ideas they might be afraid to express otherwise'', while nearly half of them believe that ``allowing commenters to remain anonymous raises the level of disrespect''.
Nevertheless, Stroud et al.~\cite{stroud}, through a survey with more than 12,000 Internet users, argue that anonymity might actually not play much of a role in uncivil discourse from commentators on the Web.
On this, Li et al.~\cite{li2017trollspot} propose a methodology to identify malicious users on commenting platforms, with an overall classification accuracy of almost 81\%.

Kalogeropoulos et al.~\cite{kalogeropoulos2017shares} show that political partisans are more likely than non-partisans to engage in commenting on news stories in social media, while Park et al.~\cite{park2011politics} reveal that it is possible to automatically predict the political orientation of news stories through the analysis of the behavior of individual commentators.
Specifically regarding comment behavior in YouTube, Ksiazek, Peer and Lessard~\cite{ksiazek2016user} explore the relationship between popularity and interaction in news videos published to this service, concluding that ``users engage with content in various ways and at differing levels, ranging from exposure to recommendation to interactivity''.

\section{Conclusions and future work}
\label{sec:conclusions}
In this paper, we present an investigation regarding comments and video content in a set of right-wing YouTube channels and compare it to a set of baseline channels.
We perform a three-layered analysis through which we examine lexicon, topics and  discriminatory bias in videos and comments from the collected channels.
\paragraph{Findings}
The two research questions proposed in Section~\ref{sec:intro} are partially answered by our analyses.
Our main findings suggest that right-wing channels are more specific in their content, discussing topics such as terrorism and war, and also present a higher percentage of negative word categories, such as \textit{agression} and \textit{violence}, while the baseline channels are more general in their topics and use more positive words. Although not capturing a difference of bias against immigrants and LGBT people, we were able to capture a negative bias against the Muslim community. When comparing comments and video hosts, we observe that, while there is a difference on the actual semantic fields, both commentators and hosts use negative words. By analyzing the implicit bias, the differences for baseline channels are not very strong, while for right-wing channels we notice a higher bias against immigrants and Muslims among captions, and a higher bias against LGBT people among comments. These findings contribute to a better understanding of the behavior of general and right-wing YouTube users.

The method presented in this study, which uses only open source tools, combines together three already established analytical procedures.
By performing these different but complementary analyses in our dataset, we are able to tackle the examined issues by distinct angles and to observe aspects that would have been ignored in one-layered investigations.
For example, lexical and topic analysis measure the presence of words that semantically convey feelings, but they are not good estimators about towards whom or what those feelings are about. Related works often use part-of-speech tagging and named entity recognition~\cite{fast2016shirtless} to tackle this problem. However,
the Word Embedding Association Test (WEAT) takes advantage of word embeddings in which words that share common contexts are located in close proximity to one another. Through this method, it is possible to measure implicit associations and then complement the lexical and topic analyses.

\paragraph{Future work}





Here, we do not handle with negation, i.e. we do not consider whether a hateful word is accompanied by a negation that reverses its meaning.
This is especially important for our lexical analysis, that simply counts the occurrence of words in given semantic fields.
The use of our multi-layered approach mitigates this problem, but, in future work, we plan to improve our analyses in this regard.
Also, we analyze our data from a synchronic point of view -- that is, we observe it as one single point in time. 
In the following steps, we plan to incorporate a temporal aspect to our investigations, since we believe that diachronic information will make it possible to elucidate to what extent do violent and discriminatory behavior in videos stimulate violent and discriminatory behavior in comments and vice versa. The incorporation of time analysis may also improve our LDA results, since it would be possible to create the notion of conversation sessions and to split the large documents that aggregate all videos' comments into smaller document sessions.

\section*{Acknowledgments}

This work was partially supported by CNPq, CAPES, FAPEMIG and the projects InWeb, MASWEB, Atmosphere and INCT-Cyber.

We would like to thank Nikki Bourassa, Ryan Budish, Amar Ashar and Robert Faris, from the Berkman Klein Center for Internet \& Society at Harvard University,
for their insightful discussions and suggestions.

\bibliographystyle{ACM-Reference-Format}
\bibliography{ref}

\end{document}